Original Paper

Dylan Hamitouche[1,2]; Youcef Barkat[1,3], BSc; Deven Parekh[1], MSc; Eva Hammer[1,4]; David Benrimoh[1,5], MD, MSc, MSc

[1] Douglas Research Center
[2] Department of Medicine, McGill University
[3] Department of Biochemistry, University of Montreal
[4] Department of Psychology, McGill University
[5] Department of Psychiatry, McGill University

## Authors' Contributions:

DH helped conceive the study, performed the analyses, and wrote and revised the manuscript.
YB helped complete the analyses and reviewed the manuscript
DP, EH reviewed the manuscript
DB: helped conceive the study and provided supervision

# Dynamic Indicators of Adherence and Retention in Digital Health Studies: A Longitudinal Observational Analysis from the Brighten Study

## Abstract

**Background:** Making optimal use of mobile health technologies requires the validation of digital biomarkers, which demands high levels of participant adherence and retention. However, current remote digital health studies have high attrition rates and low participant adherence, which may introduce bias and limit the generalizability of the findings.
**Objective:** The objective of the study is to identify longitudinal indicators of participant retention and adherence, which may serve to develop strategies to improve data collection in digital health studies and improve understanding of how study cohorts are shaped by participant withdrawal and non-adherence.
**Methods:** We performed secondary analyses on the Brighten study, which consisted of two remote, smartphone-based randomized controlled trials evaluating mobile apps for depression treatment, enrolling 2,193 participants in total. Participants were asked, after baseline assessment, to complete 7 digital questionnaires regularly. We assessed adherence to digital questionnaires, engagement (post-baseline participation), and retention rates (proportion of participants who continue completing questionnaires over time) as outcomes. We investigated the relationship between these outcomes and both static measures (e.g., demographics, average questionnaire scores) and dynamic measures (e.g., changes in questionnaire scores over time).

**Results:** The study included 2201 participants, of whom 1093 completed at least one non-baseline questionnaire, with a median completion rate of 37.6%. We found significantly higher adherence rates in participants who were less depressed on average over the course of the study (t-score=-5.63; *P*<.001) and in those who perceived clinical improvement (t-score=3.78; *P*=.001). There were significant demographic differences in adherence and engagement, specifically between gender, race, education, income, and income satisfaction. Participants who were more depressed at baseline were more likely to withdraw before completing any non-baseline questionnaire (t-score=-2.53, *P*=.01). However, participants who showed improvement in depressive symptoms during the study showed better adherence (Mann-Whitney U=127,084; *P*<.001) and retention (HR=0.78, 95% CI: [0.67, 0.91], *P*=.002), despite showing greater depressive symptoms at baseline.
**Conclusions:** We show that the clinical trajectory of participants regarding their depressive symptoms, as well as their own perception of their improvement, are important indicators of engagement, adherence, and retention. Expanding knowledge regarding these longitudinal indicators may improve interpretation of outcomes and help build strategies to improve retention and adherence in future clinical trials.

**Keywords:** mHealth; digital health; mobile health; digital biomarkers; adherence; adherence; depression; smartphone; digital interventions

## Introduction

In the past decade, researchers have dedicated significant efforts to identifying clinically useful biomarkers for preventing, diagnosing, and treating psychiatric disorders. Digital biomarkers have garnered considerable attention, as they enable the observation of long-term patterns and trends outside of the hospital or clinic environment, potentially enhancing our understanding of the course of mental illnesses.[1,2] Identifying reliable digital biomarkers has the potential to significantly improve the accuracy of diagnoses, predictions of clinical outcomes (e.g. suicidal ideation[3]) and treatment decision making regarding psychiatric disorders[4–6]. Furthermore, collecting digital biomarkers is cost-effective[7,8] and allows the possibility to provide patients with feedback through electronic reports, giving them ownership of their own medical data.[9–12]

One major challenge for the effectiveness of mobile health technologies is the identification and validation of digital biomarkers[13–16]. This process requires high levels of participant adherence and retention to ensure sufficient data is collected for validation purposes. Active data collection methods, like digital questionnaires which are often needed to help validate or contextualize digital biomarkers, intensify this challenge as they demand more time and effort than passive data collection methods such as actigraphy.[17,18] While passive measures may be useful, their validation depends on access to actively collected assessment of patient status. Therefore, understanding the drivers of participant adherence to data collection methods requiring active participation is crucial for the effective design and validation of digital biomarker collection platforms. Eysenbach described the "law of attrition", the tendency for a significant proportion of users to drop out or stop using eHealth applications before completing a trial.[19–23] This dropout may shape the cohort over time in a way that does not represent the initially recruited study population in terms of demographic features, symptomatology, and clinical outcomes, therefore inducing bias in the analyses.[16,24]

In studies requiring active remote digital participation such as completing questionnaires, a few static parameters (i.e. that do not change over the course of the study) have been associated with adherence rates (the degree to which participants follow the study protocol and complete assessments), engagement rates (the proportion of participants who engage in the study after baseline, i.e. who complete at least one non-baseline questionnaire), and retention rates (the proportion of participants who continue to complete questionnaires over time). Previous studies suggest that participants who are more depressed at baseline, younger, and less educated tend to have lower assessment adherence rates.[16,25,26] Retention rates can also vary between treatment arms involving mobile apps versus control groups, where control groups with lower engagement requirements may often show better retention, and between ethnic groups.[27] A lower annual income also tends to be significantly associated with poorer study engagement.[26]

These static parameters, often of a demographic or symptomatologic nature and considered as averages rather than longitudinal measures, provide meaningful insight into the question of participant engagement in active digital phenotyping. However, there has been less exploration of how dynamic parameters, such as longitudinal symptom change, might affect engagement. Therefore, the objective of this study was to identify novel predictive longitudinal markers of adherence, engagement, and retention to regular self-rated assessments in participants enrolling in a remote digital health study requiring active participation. To do so, we conducted secondary analyses on the Brighten study, a large longitudinal digital intervention study.

## Methods

### Study Design

The Brighten study consists of two completely remote, smartphone-based randomized controlled trials assessing the efficiency of mobile apps for treating depression. The study recruited 2,193 participants from the United States through online advertisements, with recruitment for the first version (V1) beginning in August 2014 (with recruitment period lasting 5 months) and the second version (V2) starting in August 2016 (with recruitment period lasting 7 months), enrolling 1,110 and 1,083 participants, respectively.[25,26]Participants were included in the Brighten study if they had a score of 5 or higher on PHQ-9 or a score of 2 or greater on PHQ item 10.[28] The study evaluated three interventions over a 12-week period: 1) iProblemSolve (iPST), an app designed for problem-solving therapy[29,30]; 2) Project: EVO, a therapeutic video game; and 3) Health Tips, an app offering strategies to improve mood. The group using Health Tips was considered as control group. In those two RCTs, various questionnaires were sent to participants via smartphone notifications. Most assessments were submitted either daily, weekly, or biweekly, but some questionnaires were only sent as a baseline assessment. The Brighten study also collected passive digital communication data from the V1 and V2 cohorts, along with passive mobility features from the V2 cohort. While V2 followed a protocol very similar to V1 (same interventions and active data collection methods), it specifically aimed to increase participation among individuals of Hispanic/Latino ethnic backgrounds to assess the feasibility of using digital mental health tools in this population.[28]

## Intervention apps

The iPST app implements a 7-step problem-solving therapy model to manage mood: participants select a goal and are guided through a structured action plan. Participants assigned to this app were instructed to use it at least once per week, consistent with typical clinical use.[31]

Participants assigned to the app Project: EVO (Akili Interactive Labs, Larkspur, CA, USA) were encouraged to use it six times per week for ~30 minutes per day. This video-game–based app aimed at enhancing cognitive skills related to depression using adaptive algorithms to adjust difficulty to the user's proficiency.[32] Previous studies of the app's predecessor showed that this dosage improved cognitive control and depressive symptoms in older adults.[33,34]

The control group had an app providing daily health tips, such as self-care (e.g., showering) or physical activity (e.g., walking). Participants in this group were not required to act on the tips, serving as a supportive control condition.

Importantly, all participants, regardless of their assigned intervention, completed their self-rated questionnaires in a separate study app. Our analyses specifically focused on adherence, engagement, and retention with these assessments.

## Measures

Demographics such as gender, education level, working status, income satisfaction, last-year income, marital status, race and age, as well as parameters regarding study involvement (how they heard about the study, device used, and study version) are shown in *Multimedia Appendix 1*. These parameters, with baseline PHQ-9 results, were imputed using MissForest, a non-parametric missing value imputation for mixed-type data, if less than 30% of the data was missing for that specific metric. This threshold was chosen based on validation studies demonstrating the reliability of MissForest under these conditions.[35] Participants were sent 7 questionnaires to complete at a predefined frequency specific to each questionnaire: Patient Health

Questionnaires (PHQ-9[36], weekly for the first 4 weeks and then biweekly; PHQ-2[37], daily), a 3-item sleep assessment assessing sleep onset latency, sleep duration, and time awake at night (weekly), Sheehan Disability Scale (SDS[38], weekly for the first 4 weeks and then biweekly), Patients' Global Impression of Change Scale (PGIC[39] , weekly), a survey inquiring the use of mental health services (weekly), a survey inquiring study app satisfaction (deployed at week 4,8,12), and a survey inquiring the use of other health-related apps (deployed at week 1,4,8,12).[28] All questionnaires included in our analyses can be found in Multimedia Appendix 1.

## Outcomes

We evaluated 3 key outcomes: adherence, engagement, and retention. To evaluate individual adherence, each participant's average completion rate was calculated as the ratio of completed questionnaires to the total number they were expected to complete. Engagement was binary: participants were considered engaged if they completed at least one non-baseline assessment and disengaged otherwise. Therefore, disengaged participants, by definition, had a completion rate of 0%. Engaged participants were further divided into high- and low-completion, based on the median average completion rate. Based on previous research, engaged participants were also categorized into "improvers" and "non-improvers" based on the difference between their baseline PHQ-9 score and their latest PHQ-9 score, with a minimum clinically important difference (MCID) set at 5 points.[40] Therefore, engagement captures a binary distinction (completion beyond baseline), adherence represents a continuous measure of completion, and the subgrouping serves only as a descriptive stratification of engaged participants. Participant retention was determined as the duration for which participants continued completing a specific representative questionnaire (e.g., PHQ-9) before they withdrew or stopped responding. It is important to consider that participants in the *iPST* and Project: EVO intervention groups were asked to use the apps as part of their treatment, but data on their engagement or adherence to the interventions were not available, so treatment adherence was not included as an outcome.

## Statistical Methods

To compare engagement and average completion rates across demographic groups, Mann-Whitney U tests were used for two-group comparisons, while Kruskal-Wallis tests followed by Dunn's post-hoc tests were applied for comparisons involving more than two groups.

A t-test was conducted on the average questionnaire scores and baseline scores between high and low completion groups, evaluating the importance of symptom severity as a static indicator of adherence. A t-test was also conducted on baseline PHQ-9 score between the engaged group and the disengaged group to assess if engagement was associated with baseline depressive symptoms.

To assess the potential of symptomatology as a dynamic predictor of adherence, trends in questionnaire scores over time between high and low completion groups were investigated using a Repeated Measures Generalized Linear Model (GLM).

Elastic Net Regression with SHAP values[41] was used to identify the top five predictive features for average completion rate, engagement status (engaged/disengaged), and completion group (high/low), incorporating demographic parameters, average questionnaire scores, and depressive symptom improvement. The models were subsequently retrained without including the study versions (V1/V2) as features to assess their impact. The model's generalizability was assessed by training on passive data from the V1 cohort and testing on the V2 cohort.

To study the impact of depression symptom improvement on adherence, differences in average completion rate and PHQ-9 completion rate between improvement groups were assessed using Mann-Whitney U tests. A Robust Linear Model (RLM) regression was performed with improvement status, while controlling for baseline PHQ-9 scores, to predict average completion rate. The RLM was chosen over other regression techniques to handle potential outliers in the data.[42]

Cox proportional hazards models were constructed to predict retention, measured as time to last PHQ-9 questionnaire completion, between improvers and non-improvers, incorporating demographic parameters and average scores from other questionnaires as covariates to evaluate their utility in predicting retention. We chose to study retention for the PHQ-9 rather than globally because completion rates and the frequency with which the questionnaires were sent to participants varied significantly between questionnaires, potentially compromising the reliability and validity of a comprehensive retention measure for survival analysis; the PHQ-9 was chosen as a representative questionnaire given its utility in usual clinical practice. Kaplan-Meier curves were plotted for these models to illustrate assessment retention over time.

## Results

### Outcomes

There were 2201 participants in the dataset. Among them, 1093 were engaged, meaning they completed at least one non-baseline questionnaire, and 1108 were disengaged. In the engaged group, the median average completion rate was 37.6% and the mean was 40.6% (SD=24.2%). In the engaged group, 548 participants had a mean average completion rate higher than the median and therefore belonged to the high completion group (mean average completion rate=62.3%; SD=11.5%), while 550 participants belonged to the low completion group (mean average completion rate=19.0%; SD=9.9%). The average completion rate was 20.2% (SD=26.5%) for the whole cohort.

## Demographics and study parameters

Kruskal-Wallis tests and Mann-Whitney U tests showed significant disparities in average completion rate and engagement between groups defined by gender, education, income satisfaction, last year income, race, device used, study arm, and study version (*Table 1*). The findings on average completion rate align with previous literature[16,25,26], while the engagement findings are novel, with engagement defined as whether participants chose to remain in the study after completing the baseline assessments.

**Table 1.** Significant differences ($P$<.05) in engagement rates between groups defined by demographics or study parameters.[a,b]

| | Category | Engaged (N=1093) | Disengaged (N=1108) | *P* value |
|---|---|---|---|---|
| **Gender** | | | | |
| | Male | 236 (21.59%) | 311 (28.07%) | .025 |
| | Female | 857 (78.41%) | 797 (71.93%) | .025 |
| **Education** | | | | |
| | Graduate Degree | 216 (19.76%) | 132 (11.91%) | 2.93E-05 |
| **Income** | | | | |
| | < $20,000 | 279 (25.53%) | 391 (35.29%) | 3.94E-05 |
| | 60,000-80,000 | 277 (25.34%) | 172 (15.52%) | 7.13E-07 |
| **Income satisfaction** | | | | |
| | Can't make ends meet | 769 (70.36%) | 598 (53.97%) | 1.58E-13 |
| | Have enough to get along | 194 (17.75%) | 360 (32.49%) | 1.15E-13 |
| **Race** | | | | |
| | Asian | 95 (8.69%) | 54 (4.87%) | .024 |
| | Non-Hispanic White | 640 (58.55%) | 524 (47.29%) | 7.32E-06 |
| | Hispanic/Latino | 191 (17.47%) | 366 (33.03%) | 3.43E-15 |
| **Study arm** | | | | |
| | iPST | 252 (23.06%) | 77 (6.95%) | 2.83E-24 |
| | Project: EVO | 354 (32.39%) | 181 (16.34%) | 1.24E-16 |
| | HealthTips | 369 (33.76%) | 151 (13.63%) | 8.71E-27 |
| **Study** | | | | |
| | Brighten-v1 | 742 (67.89%) | 376 (33.94%) | 3.72E-55 |
| | Brighten-v2 | 351 (32.11%) | 732 (66.06%) | 3.72E-55 |

[a] Percentages for a given category were calculated on the total number of engaged or disengaged participants. A participant is defined as being “engaged” if they completed at least one non-baseline questionnaire of any kind. If they dropped-out of the study right after baseline, without completing any non-baseline questionnaire, they are qualified as “disengaged”.

[b]Bonferroni correction applied

### Symptomatology

To evaluate the impact of baseline symptomatology and average symptom severity over the course of the study on adherence, we compared baseline questionnaire scores and average questionnaire scores between high and low completion groups. After Bonferroni correction, the mean scores of the PHQ-2, PHQ-9, and PGIC questionnaires collected over the 12-week duration of the study were significantly different between high and low completion groups (*P*<.001 for PHQ-2 and PHQ-9, *P*=.002 for PGIC). In fact, participants with lower PHQ-2 scores and PHQ-9 scores (i.e. less depressed participants), and participants with higher PGIC scores (i.e. participants who perceived that they had better clinical improvement since the beginning of the study) were more likely to be in the high completion group. This suggests that participants with lower depression levels and a perceived improvement in their clinical condition showed greater adherence compared to those with higher depression levels and a perceived worsening of their condition (*Table 2*). Mean baseline PHQ-9 score was 13.9 (SD=4.8) for the engaged group and 14.4 (SD=5.0) for the disengaged group, with the difference being statistically significant (t-score=-2.53, *P*=.01), indicating that participants who were less depressed at baseline were more likely to complete at least one non-baseline questionnaire; however, these baseline scores were not predictive of adherence in subsequent assessments.

**Table 2.** Comparisons of questionnaire scores averaged over the 12-week study period between high and low completion groups[a-g]

| | | High completion rate (SD) | Low completion rate (SD) | t-score | *P* value |
|---|---|---|---|---|---|
| **Mean Score** | | | | | |
| | PHQ-2 | 4.50 (1.52) | 5.20 (1.84) | -6.793 | 1.49E-10 |
| | PHQ-9 | 9.30 (4.87) | 11.27 (5.43) | -5.632 | 2.01E-07 |
| | Sleep Assessment | 7.23 (1.68) | 7.43 (2.09) | -1.625 | .837 |
| | SDS | 21.32 (8.50) | 22.80 (9.21) | -2.658 | .064 |
| | PGIC | 2.66 (0.97) | 2.40 (1.07) | 3.780 | .001 |
| **Baseline Score** | | | | | |
| | PHQ-9 | 13.75 (4.83) | 13.96 (4.86) | -0.709 | 1 |
| | ALC | 3.50 (2.61) | 3.64 (2.60) | -0.757 | 1 |
| | GAD-7 | 11.26 (5.78) | 11.87 (5.66) | -1.529 | 1 |

[a]Bonferroni correction applied
[b]PHQ-2: Patient Health Questionnaire-2
[c]PHQ-9: Patient Health Questionnaire-9
[d]SDS: Sheehan Disability Scale
[e]PGIC: Patients' Global Impression of Change
[f]ALC: Alcohol Use Questionnaire
[g]GAD-7: Generalized Anxiety Disorder 7-Item Scale

## Clinical Improvement

To investigate the longitudinal relationship between questionnaire scores and adherence, we conducted a Repeated Measures GLM model of mean score over time for both high and low completion groups. We found that there was significant improvement of scores on the PHQ-2 and PHQ-9 scores over time for both high and low completion rate groups ($P<.001$). Averages of PHQ-9 score for each weekly (or biweekly) assessment over time were significantly different between the two groups ($P=.021$), but the longitudinal trend was similar, with PHQ-9 scores decreasing over time. Daily PHQ-2 scores were not significantly different between the two groups, but the effect of time on the scores was statistically significant, with participants from the high completion group having a greater decrease in score over time ($P=.001$). Furthermore, the decrease in sample size regarding completion of PHQ-9 and PHQ-2 was significantly different between high and low completion groups ($P_{PHQ-9}=.02$; $P_{PHQ-2}<.001$), with the sample size from the low completion group decreasing more rapidly. For other questionnaires such as the sleep assessment, PGIC, and SDS, there was no statistically significant difference in the rate of sample size decrease between the high and low completion groups. There were no significant differences between SDS, PGIC, and sleep assessment scores between the two groups, although SDS scores decreased significantly over time for the two groups in a similar fashion ($P<.001$). The effect of time on sleep assessment scores was found to be different between the groups, but it was non-significant ($P=.08$). (*Figure 1*).

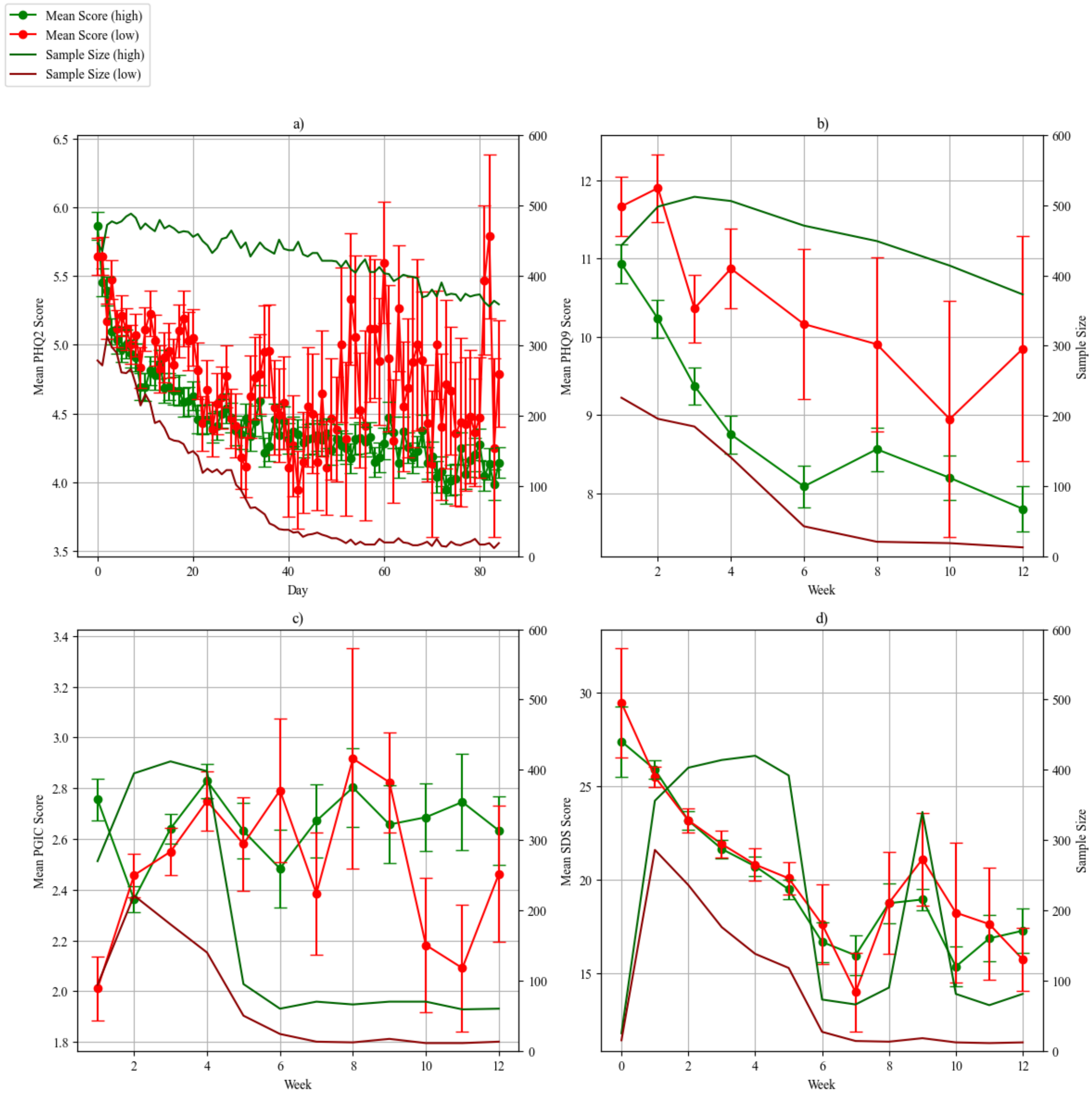

***Figure 1.*** **Mean scores over time for high and low completion groups by assessment, including sample sizes.** a) Mean PHQ-2 score over time by completion group (Intercept: 4.92, p<0.001, 95% CI [4.79, 5.05]; Day: coeff = -0.01, p<0.001, 95% CI [-0.01, -0.01]; Day × low completion group: coeff = 0.01, p<0.001, 95% CI [0.003, 0.011]). Model fit: Deviance = 15.58, Pseudo R-squared = 0.5351. b) Mean PHQ-9 score over time by completion group (Intercept: coeff = 10.38, p<0.001, 95% CI [9.65, 11.12]; Week: coeff = -0.24, p<0.001, 95% CI [-0.35, -0.14]; Week × low completion group: coeff = 0.03, p=0.697, 95% CI [-0.12, 0.18]). Model fit: Deviance = 4.02, Pseudo R-squared = 0.9767. c) Mean PGIC score over time by completion group (Intercept: 2.61, p<0.001, 95% CI [2.32, 2.89]; Week: coeff = 0.01, p=0.69, 95% CI [-0.03, 0.05]; Week × low completion group: coeff = -0.01, p=0.75, 95% CI [-0.06, 0.05]). Model fit: Deviance = 1.14, Pseudo R-squared = 0.1240. d) Mean SDS score over time by completion group (Intercept: 24.90, p<0.001, 95% CI [22.52, 27.27]; Week: coeff = -0.84, p<0.001, 95% CI [-1.17, -0.50]; Week × low completion group: coeff = -0.01, p=0.98, 95% CI [-0.48, 0.47]). Model fit: Deviance = 117.32, Pseudo R-squared = 0.8481.

To evaluate the association between clinical improvement regarding depressive symptoms and adherence, we divided the engaged cohort into improvers and non-improvers. The categorization was based on the difference between participants' baseline PHQ-9 score and their latest PHQ-9 score using the minimal clinically important difference (MCID) of 5 points as the threshold for meaningful improvement.[40] There were 410 "improvers" and 513 "non-improvers". Between those groups, there was a significant difference in average completion rate (improvers: Mean=50.7%, Median=55.5%, SD=20.6%; non-improvers: Mean=42.6%, Median=39.9%, SD=22.7%; Mann-Whitney U=127,084; *P*<.001), PHQ-9 completion rate, and baseline PHQ-9 score (*P* < .001 for all comparisons), with "improvers" having higher baseline PHQ-9 scores than "non-improvers" (Mann-Whitney U=137,574; *P*<.001). Mean baseline PHQ-9 score was 15.3 (SD=4.7) for improvers and 12.6 (SD=4.6) for non-improvers. The RLM Regression model showed that the average completion rate of "non-improvers" was, on average, 8.85% (*P*<.001) lower than "improvers". However, for both groups, baseline PHQ-9 scores were not significantly associated with average completion rates (*P*=.06). (*Figure 2*).

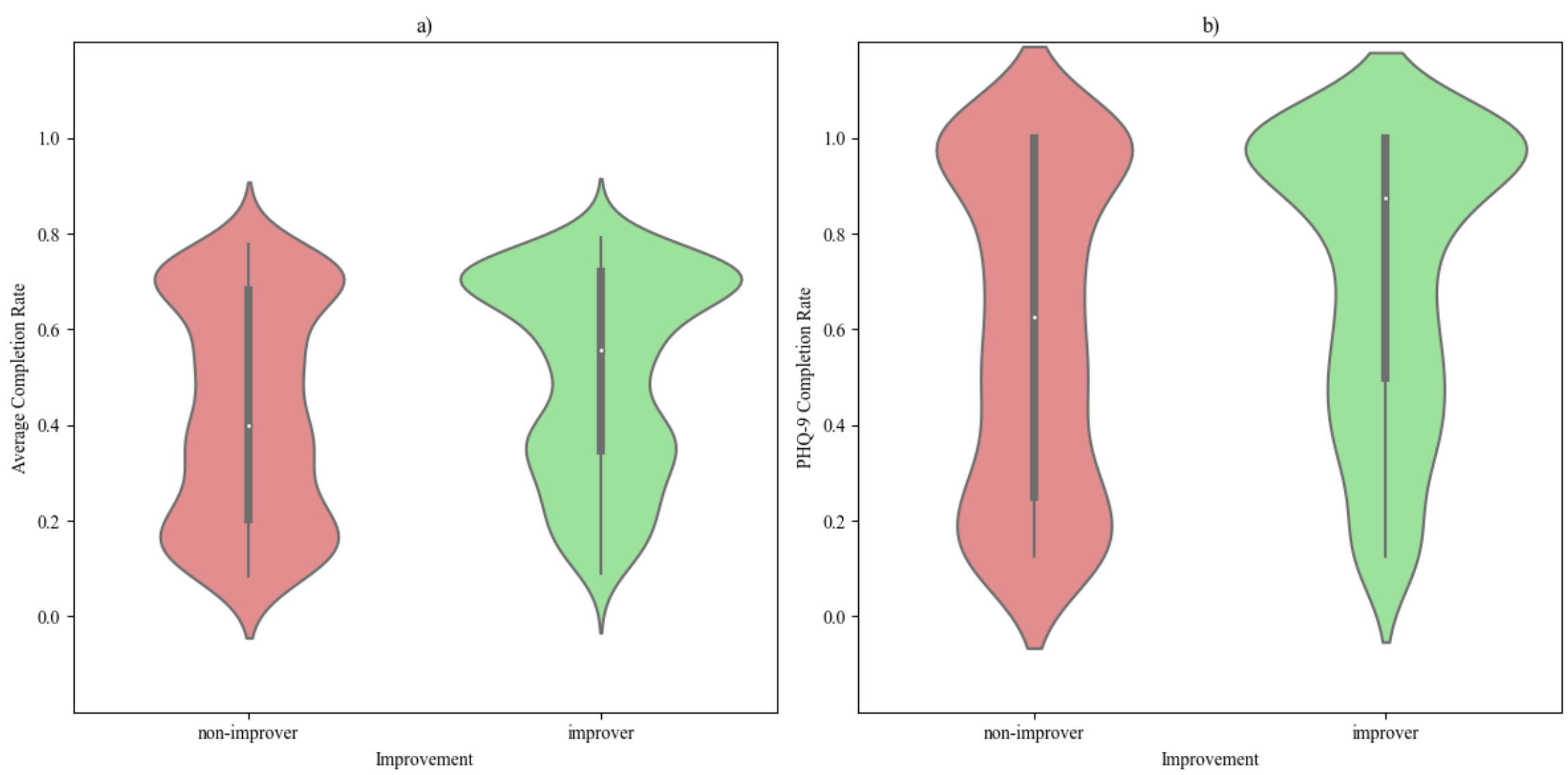

***Figure 2.*** **Comparison of adherence between improvers and non-improvers.** a) Average completion rate (Mann-Whitney U score = 127084; p-value < 0.0001 b) PHQ-9 completion rate (Mann-Whitney U score = 125418; p-value < 0.0001). Average completion rate is defined as the total number of questionnaires completed divided by the total number of questionnaires that were sent to participants. PHQ-9 completion rate is defined as the total number of PHQ-9 questionnaires completed over the total number of PHQ-9 questionnaires that were sent to participants.

To study PHQ-9 retention, defined as the duration for which participants continued completing the PHQ-9, we performed a Cox proportional hazards model to compare improvement groups, while using demographic parameters, average scores from other questionnaires, and study version as covariates. The Cox proportional hazards model revealed significant differences in survival probabilities between the two groups (HR=0.78, 95% CI: [0.67, 0.91], *P*=.002), with the event being defined as the day a participant completed their last PHQ-9 questionnaire within 12 weeks from the start of the study. The median survival time was 70 days (Mean=56.8, SD=24.7) for improvers (n=332) and 49 days (Mean=49.4, SD=25.6) for non-improvers (n=329), indicating that non-improvers stopped completing PHQ-9 questionnaires 7.4 days earlier on average compared to improvers (Mann-Whitney U= 64632, *P*<.001). We found that age (*P*=.002), mean score of sleep assessment (*P*=.01), and study version (*P*<.001) were significant features for predicting retention regarding PHQ-9 completion *(Figure 3*).

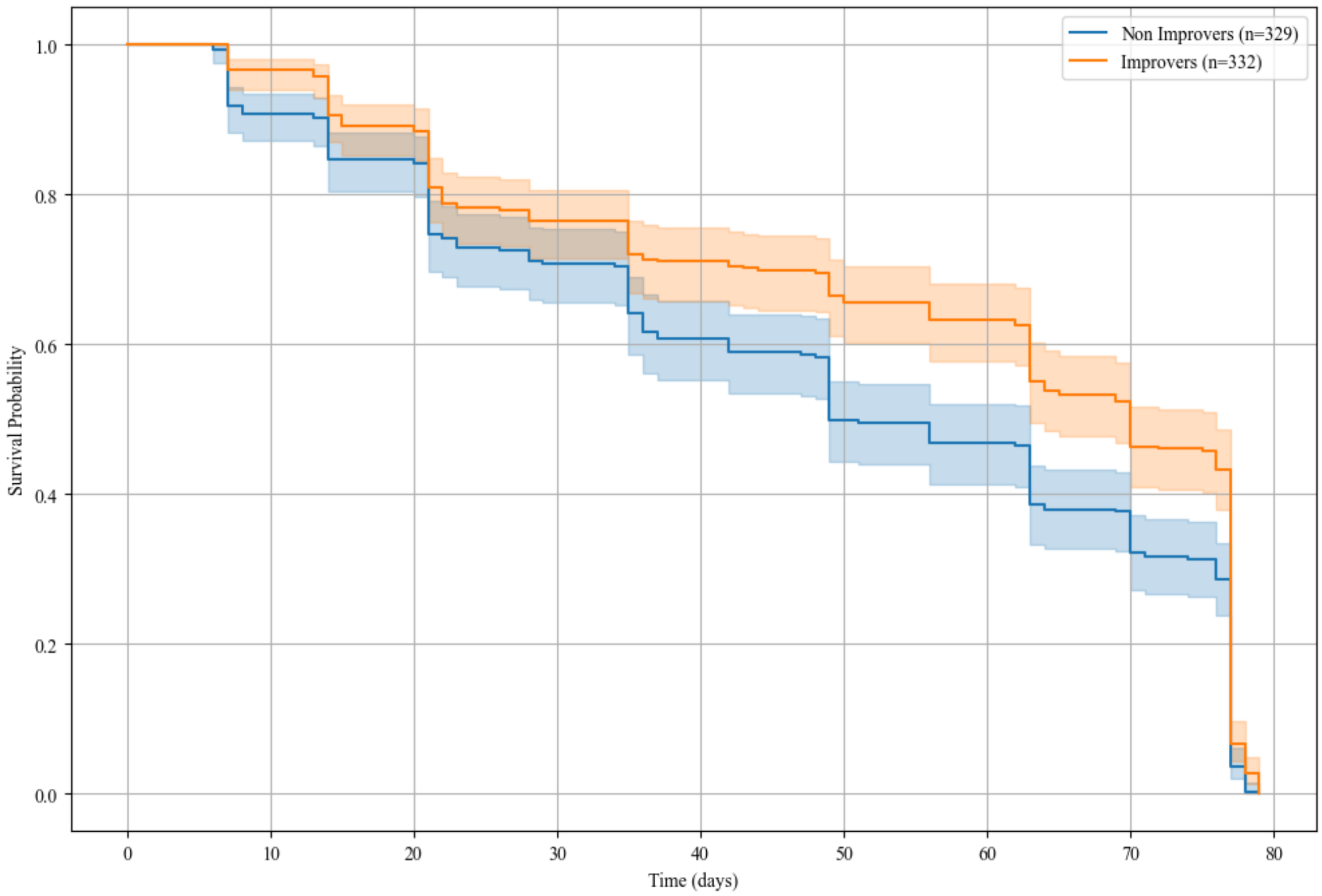

***Figure 3.*** **Kaplan Meier curves showing retention for improvers versus non-improvers, with the event defined as the last PHQ-9 questionnaire completed within the 12-week study duration (n=661).** Significant survival differences noted (Improver Coef = -0.25, Exp(coef) = 0.78, p < 0.005). Key predictors: Age: Coef = -0.01, Exp(coef) = 0.99, p < 0.005). Mean Score Sleep: Coef = 0.06, Exp(coef) = 1.06, p = 0.01). Study Brighten-v2: Coef = 0.79, Exp(coef) = 2.19, p < 0.005). Bonferroni correction applied.

### Key indicators of adherence and engagement

To assess what features were the most statistically useful in predicting average completion rate and engagement, we trained an Elastic Net Regression model on 80% of the data and tested it on the remaining 20%. SHAP values were used to assess the relative importance of each feature and evaluate their contributions to the model's predictions (*Figure 4*). For predicting average completion rate, the 5 most useful features in predicting a high average completion rate, in order of importance, were (n=701, MSE=0.04): 1) Being in the control group; 2) Having a low mean PHQ-2 score; 3) Being older; 4) Being in the V1 cohort; 5) Improving in depressive symptoms.

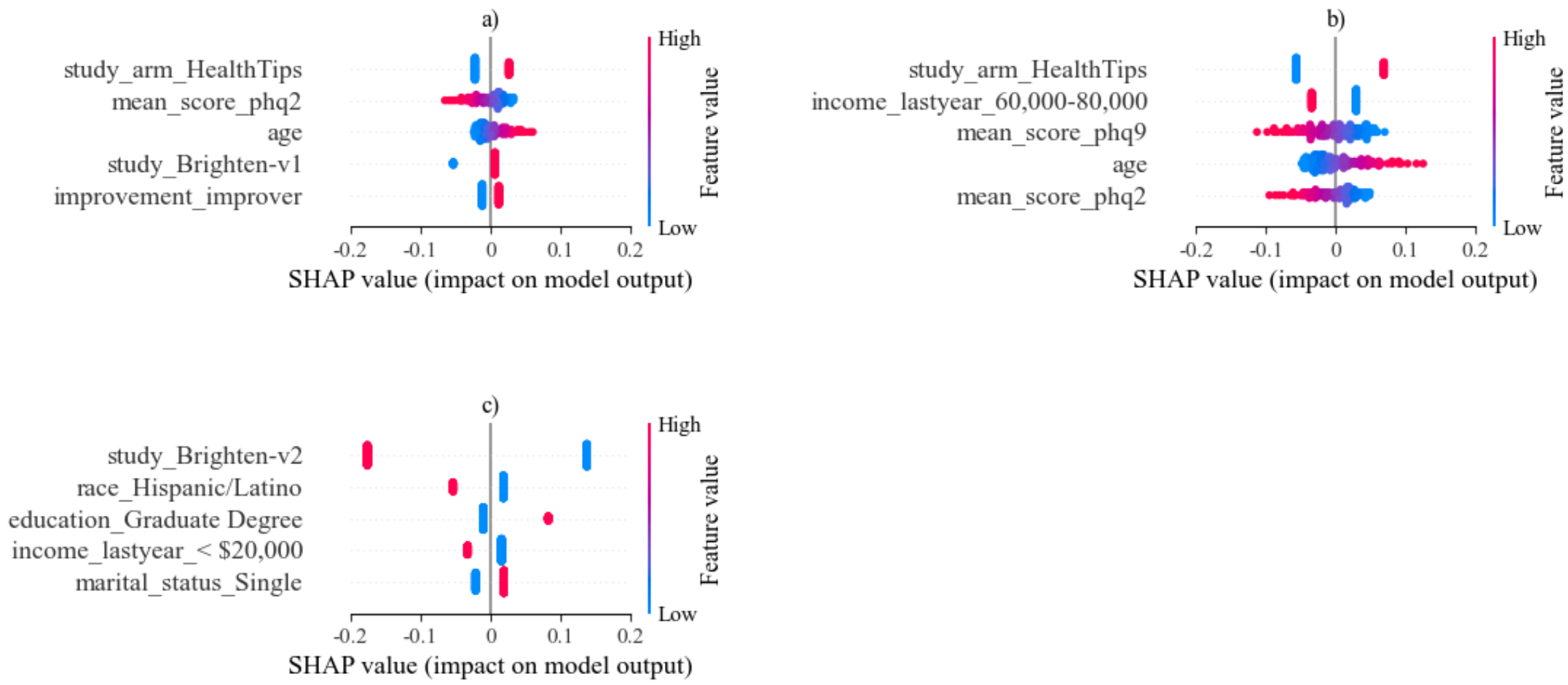

**Figure 4. Top 5 features and their SHAP value for predicting compliance metrics using Elastic Net Regression.** a) Top 5 features for predicting average completion rate (n=701, MSE=0.04). b) Top 5 features for categorizing patients into high and low completion groups (n=701, F1-score accuracy = 0.72). c) Top 5 features for predicting if a participant will be "engaged" or "disengaged" (n=2201, F1-score accuracy = 0.66).

The 5 most important features that were associated with being in the high completion group were (n=701, F1-score = 0.72): 1) Being in the control group; 2) Not being in the $60,000-$80,000 annual income group; 3) Having a low mean PHQ-9 score; 4) Being older; 5) Having a low mean PHQ-2 score.

The 5 most important features that were associated with "engaging" in the study were (n=2201, F1-score = 0.65): 1) Being in the V1 cohort; 2) Not being a Hispanic/Latino person; 3) Having a graduate degree; 4) Having an annual income greater than $20,000; 5) Being single.

To assess the impact of study version (V1/V2) on the models' predictability, we conducted an exploratory analysis where we excluded it from the features and found that the categorization models (engagement and completion group) performed slightly worse, but the 5 most important features were generally similar (F1-score decreased from 0.72 to 0.67 for completion group categorization, and from 0.66 to 0.60 for engagement categorization). The average completion rate model MSE did not change, although the $R^2$ decreased from 0.16 to 0.13.

When trained on the V1 cohort and tested on the V2 cohort, the categorization models performed poorly (F1-score = 0.32 for engagement categorization, F1-score = 0.46 for completion group categorization). The model predicting average completion rate had slightly worse predictive performance (MSE=0.04; $R^2$ Score = 0.12).

## Discussion

### Key findings

We found that subjective perception of a participant's own clinical trajectory (PGIC score) was associated with assessment adherence, with participants having a perceived improvement in their clinical condition showing greater adherence. We also found that increased depressive symptoms at baseline were not significantly associated with adherence, but rather with engagement, thus suggesting that more depressed participants were more likely to withdraw right after baseline assessment. Furthermore, participants who improved regarding depressive symptoms had greater assessment adherence and retention, despite having greater depressive symptoms at baseline. Previous research has established a link between baseline depressive symptoms and engagement[13]. This study makes novel contributions by exploring the associations between perceived clinical improvement and adherence, as well as the connection between depressive symptom improvement, baseline depression severity, retention, and adherence.

Despite offering monetary incentives, the researchers conducting the Brighten study observed poor engagement and adherence, highlighting the need to investigate this issue further and explore innovative strategies to address this limitation.[25] The demographic parameters we found associated with lower adherence and engagement are similar to what has been found in previous literature (i.e. male gender, lower education, lower income, poorer income satisfaction, and Latino/Hispanic race were associated with poorer adherence)[16,25,26]. However, we found that a perceived worsening of clinical condition throughout the study was associated with less assessment adherence, suggesting that a participant's own perception of their change in activity, symptoms, emotion and quality of life can influence their willingness to complete assessments.

Regarding completion of the PHQ-9 questionnaire, there was a significant difference in retention probability, with improvers in terms of depressive symptoms having more chance of continuing to complete assessments than non-improvers. Age, mean score of sleep assessment, and study version were significant features for predicting retention. Therefore, the global change in depressive symptoms over time was associated with study retention and adherence rate. These findings highlight the possibility of attrition bias in digital health studies, leading to an overestimation of symptom severity improvement. Furthermore, this kind of bias could lead to misleading conclusions about the effectiveness of treatments and affect clinical decision-making, reinforcing the importance of controlling for attrition rates in digital health studies.[43] However, a possible limitation of the retention analyses is the lack of insight regarding improvement for non-improvers who stopped completing answering questionnaires early in the study, as they may have eventually experienced symptom improvement that went unrecorded.

Study arm was the most important indicator of assessment completion rate, with participants from the control group having higher completion rates than participants using the Project: EVO app or the iPST app. Moreover, the finding that participants in the iPST group, who were encouraged to use the app once per week, showed higher adherence than those in the Project: EVO group, who were asked to use it six times per week, suggests a trade-off between effort and study engagement. Participants required to invest more time in the digital intervention were more likely to skip assessments. We found that baseline PHQ-9 score, although it was predictive of engagement (serving as an initial "filter" for participants continuing in the study), was not an important indicator of adherence, highlighting the need to focus on clinical trajectories in addition to static measures. Additionally, participants who improved significantly in their depressive symptoms had higher completion rates than participants who did not, and their baseline PHQ-9 scores were significantly higher, although, as mentioned, baseline PHQ-9 score is not a direct indicator of completion rate. This suggests that while initial severity is not directly linked to longitudinal adherence, positive clinical progress may encourage consistent assessment completion. In fact, we found that participants who significantly improve regarding depressive symptoms are less likely to withdraw from active completion of assessments than participants who do not. Interestingly, it is also worth considering that the capacity to regularly complete questionnaires might be a potential predictor of future improvement in depressive symptoms.

### Limitations

This study presents several limitations. The most important limitation is the lack of data on participants' use of the intervention apps, particularly the Project: EVO group, whose participants were encouraged to use the app 6 times per week. As a result, we were unable to assess whether engagement with the intervention apps influenced participants' engagement with questionnaire completion. However, our primary interest was in questionnaire completion, as this measure is more directly comparable across digital health studies and thus the most applicable for broader research contexts. Also, the GLM Repeated Measures analysis revealed intriguing patterns and trends in the effect of time and adherence on questionnaire scores, but the rapid decline in sample size over time limits the strength of any conclusions that can be drawn. The Elastic Net Regression models showed only modest performance, with clear potential for improvement. Feature engineering, collection of larger feature sets, hyperparameter tuning, and time-series transformations are potential strategies to explore that could enhance the accuracy and robustness of the models by improving their ability to recognize underlying patterns, as well as finding temporal dynamics within the data[44–46]. Given the model's limited fit, SHAP value interpretations for top features should be regarded as exploratory. Furthermore, we were not able to generalize the model's predictability across cohorts, highlighting the need for further validation with diverse datasets to ensure its applicability in a realistic trans-diagnostic population[47,48]. The limited data available for passively collected variables, such as communication and mobility features, resulted in inconclusive analyses including these variables (see Multimedia Appendix 3).

## Future Directions

To address adherence and retention challenges in future studies, implementing continuous monitoring systems to identify participants at risk of dropping out could be beneficial.[49] Targeted outreach strategies, such as personalized reminders or support interventions, may help engage these individuals and improve adherence.[50,51]Monetary incentives have also been shown to be effective in increasing adherence and retention rates, though this may be impractical when attempting more naturalistic designs.[52–54] Beyond these approaches, qualitative research is crucial to better understand the underlying reasons for disengagement. Prior user-centered studies indicate that underserved users often cite stigma, cultural mismatch, and usability issues as barriers, while individuals with lower education or digital literacy report practical challenges and a higher perceived time burden when using mHealth apps.[55,56] Conducting more extensive qualitative work would provide valuable insights into how participants perceive their improvement and clinical trajectories, thereby offering a more comprehensive understanding of user challenges and informing strategies to strengthen engagement, adherence, and retention.

## Conclusion

Although demographics are useful in predicting adherence, engagement, and retention in digital health studies, we show that dynamic measurable parameters regarding symptomatology and symptom change are also important in understanding longitudinal patterns of attrition and adherence, and that they play a crucial role in shaping the remaining cohort as long-term studies progress over time[16,24]. We have shown that patient impressions of their own improvement and the clinical trajectory of participants regarding depressive symptoms are important indicators of engagement, adherence, and retention. Further research is needed to assess the causality and implications of this effect, and to determine optimal strategies for countering the bias that this phenomenon introduces into datasets.

## Acknowledgements

The authors acknowledge financial support from the Mach-Gaensslen Foundation of Canada, the Dr. Clarke K. McLeod Memorial Scholarship, the Douglas Research Center, the Fonds de recherche du Québec – Nature et technologies (FRQNT) (STRATÉGIA grant), and the Fonds de Recherche du Québec - Santé (FRQS) (Junior 1 Grant).

## Conflicts of Interest

Dr. David Benrimoh is a founder and shareholder of Aifred Health, a digital mental health company. Aifred Health was not involved in this research.

## Multimedia Appendix 1

Significant differences in average completion rate between groups defined by demographics or study parameters, and questionnaires analyzed in the study.

## Multimedia Appendix 2

Kaplan-Meier curve showing retention of improvers versus non-improvers based on PHQ-2 scores

## Multimedia Appendix 3

Top 5 features and their SHAP value for predicting adherence metrics using Elastic Net Regression trained on demographics, study parameters, average questionnaire scores, and passive data features

## Abbreviations

ALC: Alcohol Use Questionnaire
CI: confidence interval
eHealth: electronic health
GAD-7: Generalized Anxiety Disorder 7-Item Scale
GLM: generalized linear model
HR: hazard ratio
iPST: iProblemSolve
MCID: minimum clinically important difference
MSE: Mean Squared Error
PGIC: Patients' Global Impression of Change
PHQ-2: Patient Health Questionnaire-2
PHQ-9: Patient Health Questionnaire-9
RCT: randomized controlled trial
RLM: robust linear model
SDS: Sheehan Disability Scale
SHAP: Shapley additive explanations